\documentclass[12pt]{iopart}

\usepackage{graphicx}
\begin{document}

\title{Grover's algorithm and human memory}

\author{Riccardo Franco}

\begin{abstract}
In this article we consider an experimental study showing the influence of emotion regulation
strategies on human memory performance: part of such experimental results are difficult to explain
within a classic cognitive allocation model. We provide a first attempt to build a model of human memory processes based on a
quantum algorithm, the Grover's algorithm, which allows to search a particular item within an
unsorted set of items more efficiently than any classic search algorithm. Based on Grover's
algorithm paradigm, this new memory model results to have interesting features: it is an
iterative process, it uses parallelism and interference effects. Moreover, the strength of
such interference effects depends on a parameter of the generalized Grover's algorithm, called the phase, which admits an
interpretation in terms of the emotions involved by the items and by the emotion regulation
strategies. Thus we show that a reasonable choice of the phase is able to describe correctly
the experimental results we consider.
\end{abstract}


\section{Introduction}
In a traditional computer, information is encoded in a series of bits which are manipulated
via boolean  logic gates arranged in succession to produce the end result.  Similarly, a
quantum computer manipulates qubits (the quantum analog of bits) by executing a series of
quantum gates which form a quantum algorithm.
A classical computer is effectively incapable of performing many tasks that a quantum computer
could  perform with ease. This is consistent with the fact that the simulation of a quantum
computer on a classical one is a computationally hard problem \cite{Fey}.

The quantum computation uses two important effects: the parallelism, due to the linearity of
the  quantum gates, and the interference. The combination of parallelism and interference is
what makes quantum computation powerful, and plays an important role in quantum algorithms
\cite{Aharonov}.
The first quantum algorithm which combines interference and parallelism to solve a problem
faster than classical computers, was discovered by Deutsch et al. \cite{Deutch}. This
algorithm addresses the problem we have encountered before in connection with probabilistic
algorithms: distinguish between "constant" and "balanced" databases. The quantum algorithm
solves this problem exactly, in polynomial cost, while classical computers cannot do this, and
must release the restriction of exactness.  Shor's algorithm (1994) is a famous polynomial
quantum algorithm for factoring integers, and for finding the logarithm over a finite field
\cite{Shor}. For such problem, the best known classical algorithms are exponential.

In 1995 Grover \cite{Grover1} discovered an algorithm which searches an unsorted database of
$N$ items and finds a specific item in $\sqrt{N}$ time steps. This result is surprising,
because intuitively, one cannot search the database without going through all the items.
Grover's solution is quadratically better than any possible classical algorithms. The Grover's
algorithm is based on the hypothesis that exists a function $C(i)$ (with $i=1,..,N$) such that
$C(\overline{i})=1$ for the marked item $\overline{i}$, while $C(i)=0$ for the others. Such
function can be applied simultaneously to all the database entries, but the useful information
is extracted by repeating a suitable operation which makes use of interference. Some
generalizations of Grover's algorithm have been provided, such as \cite{Long, Long1, Grover
general}.

There are many features of quantum algorithms  that induces us to use concepts of quantum
computation in describing human memory effects. First of all, there is evidence that we retain
more than we can retrieve: the effectiveness of retrieval can depend on many factors, such for
example emotions \cite{Gross1998b, Baumeister1998}. This seems consistent with the generalized
Grover's algorithm, where the effectiveness of quantum search depends on the intensity of the
interference effects which extract the information. In other words, such algorithm is
consistent with the fact that we are potentially  able to recall the information, but we have
to extract the correct answer from the other potential outputs by using interference effects.
We provide a psychological interpretation of the important quantum gate known as
\textit{selective phase rotation} in terms of emotional influence. To confirm such
suggestions, we recall the fact that the forgetting mechanism is also connected to
interference effects. Secondly, the process of search in Grover's algorithm has an optimal
number of iterations: a lower or higher number of iterations leads to a lower effectiveness of
the search. Such effects, known respectively as undercooking and overcooking, seem to be
consistent with the fact that sometimes we do not remember because we have too less or too
much time to think.

This article is addressed both to physicists and to psychologists. To this end, we give the
mathematical details in the appendix, while in the body of the paper we introduce only the
mathematical  basic concepts which are essential to the understanding.

Our work is part of a research topic which can be called "quantum cognition", which studies
within the quantum formalism experiments of logical fallacies \cite{Rfranco_conj2,
Rfranco_inverse}, decision theory \cite{Decision} and semantic \cite{Semantic}.
%
\section{Memory and emotion regulation}
Quite recently,  researchers in cognitive science have begun to examine how individuals
regulate  their emotions and to document what consequences such attempts at emotion regulation
have \cite{Gross1998b}. It is quite natural to accept that emotions frequently arise when
important goals are at stake and thus when peak cognitive performance is critical.
If from a point of view emotion regulation is quite natural, from another it seems to be
\textit{effortful}: this means that emotion regulation can deplete mental resources. For
example, emotion-regulation participants were found to solve fewer problems than no-regulation
participants \cite{Baumeister1998}.

Two emotion-regulation strategies have been identified: the \textit{reappraisal} (for example,
appraising an upcoming task as a challenge rather than a threat, or looking at the situation
as an external observer) and the \textit{expressive suppression} (inhibition of the urge to
act on emotional impulses that press for expression). For a list of references about examples
where emotion regulation consumes cognitive resources, see \cite{Gross1998b}. A particular
object of study is the regulation of negative emotions, and its influence on subjects' memory.
In the present article we consider an experiment  performed by \cite{Gross1998b}  in order to
study how the expressive suppression and the reappraisal conditions influence the ability to
remember. In such experiment participants watched emotion-eliciting slides (distinguished
between high-level or low-level emotion eliciting slides), and answered to questions about
verbal and nonverbal memory.
%
%
%
\subsection{Description of the experiment}
Experiment 2 of \cite{Gross1998b} attempted to show that experimentally manipulating
\textit{reappraisal} and expressive suppression in a controlled laboratory setting would
differentially influence memory for information presented during the induction period.
Moreover, it studied whether suppression would lead to poorer memory even when \textit{low or
high levels of emotion} were elicited.

Eighteen slides were presented in two sets of nine slides each on a  television monitor. As
slides  were presented, three data --a name, an occupation, and a cause of injury-- were
presented using an audio recording. Slides were presented individually for 10 s;  slides
within each set were separated by 4 s. Some of the slides (9) showed people who appeared
healthy because their injuries had happened a long time ago (low-emotion slide set), but other
slides (9) showed people who appeared gravely injured because they had been photographed
shortly after sustaining their injuries (high-emotion slide set).

Participants were randomly assigned to: the watch condition (40), the expressive suppression
condition (20) and the reappraisal condition (22). Participants viewed a first set of nine
individually presented slides (low-emotion for example), and after a self-report emotion
experience measure, they viewed a second set of slides (high-emotion for example). Finally
they performed by a cued-recognition test of the slides and a cued-recall test of the  orally
presented biographical information

Cued-recognition test (non-verbal memory): participants were shown 18 photo spreads, one
corresponding to each of the 18 slides they saw in the first phase of the experiment. For each
photo spread, participants were asked to identify which of four alternatives most closely
resembled the slide they had seen earlier. The correct alternative was the same image
participants had seen earlier, with the only difference being that it was reduced in size.
Incorrect alternatives were generated by modifying particulars of the original image.
Participants had 8s to view each photo spread and to give their answer.

Cued-recall test (verbal memory): after viewing the photo spreads, participants viewed
the original slides one more time. This time, they were asked to write down
the information that had been paired with each slide during the initial
slide-viewing phase (i.e., name, occupation, injury).

Results for verbal memory: only suppression participants showed a reliable decrease
in memory (13\%), compared with watch  (18\%) and reappraise (16\%) conditions.
Overall, verbal information was remembered less well if it accompanied high-emotion slides.

Results for non-verbal memory: unexpectedly, reappraisal participants were more likely to
correctly identify high-emotion slides they had seen earlier than watch participants.
Low-emotion case:  watch (43\%), reappraise (40\%), suppress (35\%) High-emotion case: watch
(37\%), reappraise (48\%), suppress (40\%). In the following, we will focus our attention on
the results of non-verbal memory, which are quite surprising from the point of view of  a
classic allocation model: in fact, even in the high-emotion case the watch condition seems to
involve no cognitive costs in emotion regulation., and thus it should evidence the highest
success percentages. The quantum Grover's algorithm provides a completely different model,
where the emotions seem to play a key role in manipulating interference effects which drive
the memory process.
%
%
%
\section{The generalized Grover's algorithm}
Grover's algorithm, invented by Lov Grover in 1996 \cite{Grover1}, is a quantum algorithm for
searching an unsorted database.  Classically, searching an unsorted database with $N$ items
requires a linear search. This means that the time we need is in mean proportional to the
number of items $N$. On the contrary, Grover's algorithm requires a number of steps
proportional to $\sqrt{N}$, from which it is easy to note that such quantum algorithm is more
efficient than the classical one. A simple example, cited in \cite{Grover1}, is a phone
directory containing $N$ names arranged in completely random order. If we want to find
someone's phone number with a probability of $1/2$, any classical algorithm (deterministic or
probabilistic) will need to look at a minimum of $N/2$ names.

We give now a simple description of a particular generalization of Grover's algorithm which is
useful for our purposes \cite{Long, Long1}, evidencing three main features:\\
1) The preparation of the quantum register: given the $N$ items, we attribute to each item
$i=1,..,N$ the same \textit{quantum amplitude}
\begin{equation}\label{superposition}
\psi(i)=1/\sqrt{N}\,.
\end{equation}
We recall that the quantum amplitude $\psi(i)$ is a generalization of the classic probability
weight. The probability to measure any item $P(i)$ is the quare modulus of the amplitude
$P(i)=|\psi(i)|^2$ and thus $P(i)=1/N$. This means that each item is supposed to have
initially the same probability to be measured. We note that the quantum formalism allows for
more general evenly distributed superpositions. For example, the amplitude associated to a
particular item $i$ could be a complex number $e^{i\phi}/\sqrt{N}$. Since the probability to
measure such item is the square modulus of the amplitude, we would have
$|e^{i\phi}/\sqrt{N}|^2=1/N$. The complex factor, called the \textit{phase}, has an important
role in interference effects.
\\
2) The \textit{function of the marked item} $C(i)$ such that:
\begin{equation}\label{marked}
C(\overline{i})=1,\,\,C(i)=0\,\forall i\neq \overline{i}
\end{equation}
In other words, the Grover's algorithm hypothesizes the existence of a function which is true
for the marked item and false for the other items.
\\
3) The \textit{iterative search engine}, representing the quantum gate which uses the function
$C(i)$ and the interference effect to make the probability relevant to the marked item
$P(\overline{i})$ near to 1. The search engine is characterized by a specific parameter, the
phase $\phi$. The probability to find the marked item $P(\overline{i})$ after $J$ iterations
and with the characteristic phase $\phi$ is
\begin{equation}\label{prob_j}
P(\overline{i})=sin\left\{\left[2Jsin(\phi/2)+1\right]arcsin\left(\frac{1}{\sqrt{N}}\right)\right\}^2
\end{equation}
We can see from this formula that the probability to find the marked item is a periodic
function of the number of iterations $J$. To obtain $P(\overline{i})\simeq 1$, the quantum
search engine has to be repeated $J_{opt}$ times, where $J_{opt}$ is the optimal number of
iterations. In the high $N$ case we have:
\begin{equation}\label{j_opt}
J_{opt}\simeq \frac{\pi\sqrt{N}/4-1/2}{sin(\phi/2)}\,.
\end{equation}
In the high $N$ condition, a number of iterations lower than $J_{opt}$ leads to a probability
to find the marked item lower than 1: this fact is known as \textit{undercooking}. Also a
number of iterations higher than $J_{opt}$ leads to a probability to find the marked item
lower than 1: this fact is known as \textit{overcooking}.
%
%
%
\subsection{A quantum-like model for memory}
In this section a  quantum-like model for human memory is proposed, based on a generalization
of Grover's algorithm of Long et al. \cite{Long,Long1}.  Such model does not assume that the
human mind has a quantum nature, although some evidences and models have been provided by
Vitiello et al. \cite{vitiello}: it simply defines a formal mathematical model, based on a
quantum algorithm, which is able to give a good description of human memory experiments.

First  of all, we give  some remarks about cued-recognition task and cued-recall task: in the
first the subject is presented some alternatives, and the searched item is part of them. On
the contrary, in the second the subjects have to give the searched item without having any
alternatives, but only with the aid of a cue. However, in the cued-recall task the
alternatives have been previously presented during the first part of the experiment. If
follows that in both cases we will assume the existence of a quantum register, as described
before.

Thus, given a memory task over $N$ alternative items (recall or recognition), the main
hypotheses of our quantum-like model of memory are:
\\
1)  subjects first assign \textit{equal amplitudes} $\psi(i)=1/\sqrt{N}$ to the possible
items, as described in equation (\ref{superposition});
\\
2) subjects are provided a \textit{function of the marked item} as evidenced by equation
(\ref{marked}). In other words, an unconscious knowledge of the marked item is supposed;
\\
3) subjects perform an iterative process, by repeating for a  number of times $J$ the same
\textit{searching engine}, characterized by a specific phase $\phi$, as described by equation
(\ref{prob_j}), which evidences the probability to find the marked item after $J$ iterations.
\\
4) the \textit{characteristic phase} $\phi$ of a search engine is determined  by two factors:
a) it increases with the level of emotion elicited by the items or by the cues, and b) it
decreases with the emotion regulation strategy. Thus, given the characteristic phases in watch
strategy for low emotion $\phi(W_1)$ and high emotion $\phi(W_2)$, reappraisal strategy
$\phi(R_1),\phi(R_2)$ and suppression $\phi(S_1),\phi(S_2)$, we impose:
\begin{eqnarray}\label{condition1}
\phi(W_1)>\phi(R_1)>\phi(S_1)\\\label{condition2}
\phi(W_2)>\phi(R_2)>\phi(S_2).
\end{eqnarray}

Given a fixed number of iterations $J$ and a fixed number of items $N$,  the probability to
find the marked item is given by formula (\ref{prob_j}) and depends only on the parameter
$\phi$. If we associate a unit of time to each iteration, the fixed $J$ condition is
equivalent to a fixed-time memory task.

We consider the test of non-verbal memory of experiment 2 in \cite{Gross1998b}. The number of
items is given not only by the total number of images, but also by the  details which have
been modified in each spread. We estimate a value of $N=80$. In figure 1 we show the
probability to find the marked item as a function of $\phi$ for different number of iterations
$J$.
\begin{figure}[h]\label{fig1}
\centering
\includegraphics[width=13.5cm,height=8.8cm]{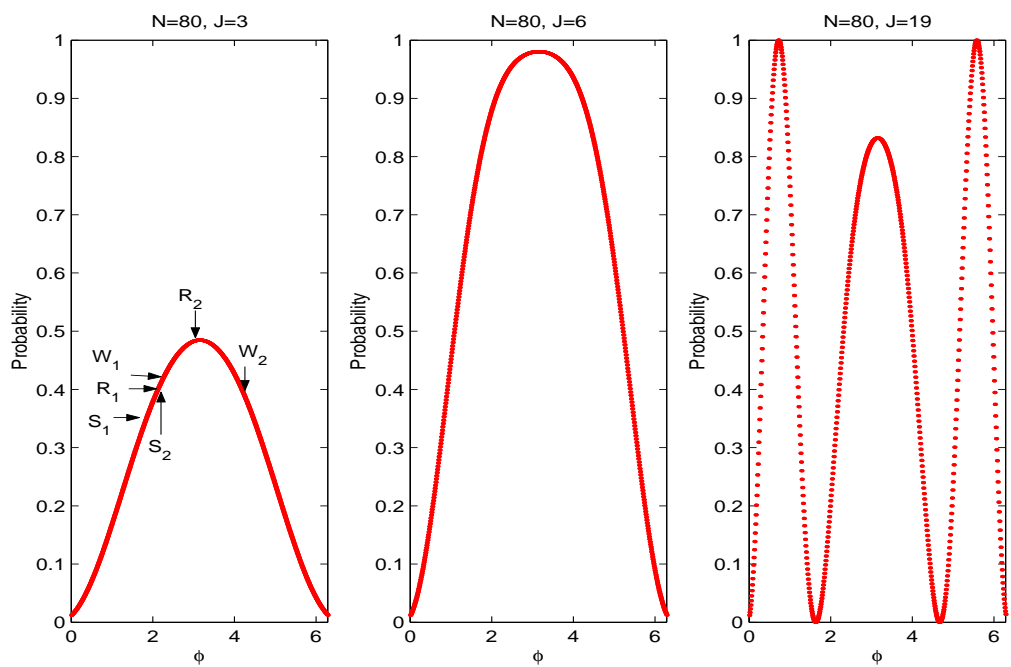}
\caption{Probability to find the marked item as a function of $\phi$ for different values of
$J$}
\end{figure}
We can see that in the first panel the probability 1 to remember  the market item is never
reached $P(\overline{i})<1$: this entails a condition of \textit{undercooking}, since the
number of  iterations is not sufficient. The second panel shows a probability near to 1 at
$\phi=\pi$, which is consistent by the original Grover's algorithm; for values different from
$\phi=\pi$, we have undercooking. The third panel evidences a central region where $P<1$
between two peaks of value 1: such region describes the \textit{overcooking} condition.

Since in experiment 2 of \cite{Gross1998b} the optimal situation of probability 1 is never
reached, this seems to imply that we are in a condition similar to the left panel of figure 1,
where $J$ is fixed to 3. Thus the number of iterations is not sufficient to completely
remember the items. In left panel of figure 1, we have evidenced the points corresponding to
the experimental situation. In particular, we call $W_1$, $R_1$ and $S_1$ the situations of
watch, reappraisal and suppression respectively for the low emotion case, while we call $W_2$,
$R_2$ and $S_2$ the situations of watch, reappraisal and suppression respectively for the high
emotion case. We attempted to evidence such phases on the left panel of figure 1: it is
important to note that  we have not at the moment a precise mathematical law to calculate
them, we only have conditions (\ref{condition1},\ref{condition2}). The situation $W_2$ results
to have a characteristic phase higher that $\pi$, and thus a sub-optimal recognition
condition. In the table below we summarize such data as the corresponding characteristic
phases.
\\\\
\begin{tabular}{|c||c|c|c|}\hline
 Level of emotion  & Watch & Reappraise &  Suppress \\\hline\hline
   Low     &
\begin{tabular}{c}
$W_1$ situation  \\
$P(\overline{i})=0.43$  \\
$\phi\simeq 2.8$
\end{tabular}
&
\begin{tabular}{c}
$R_1$ situation  \\
$P(\overline{i})=0.40$  \\
$\phi\simeq 2.5$
\end{tabular}
&
\begin{tabular}{c}
$S_1$ situation  \\
$P(\overline{i})=0.35$  \\
$\phi\simeq 2.0$
\end{tabular}
\\\hline
   High
&
\begin{tabular}{c}
$W_2$ situation  \\
$P(\overline{i})=0.37$  \\
$\phi\simeq 4.3$
\end{tabular}
&
\begin{tabular}{c}
$R_2$ situation  \\
$P(\overline{i})=0.48$  \\
$\phi\simeq 3.6$
\end{tabular}
&
\begin{tabular}{c}
$S_2$ situation  \\
$P(\overline{i})=0.40$  \\
$\phi\simeq 2.5$
\end{tabular}
\\\hline
\end{tabular}
\\\\
We stress that the numerical values of the phase, as well as the number of items $N$ are only
reasonable values, based on the experimental set and on the results.
%
\section{Conclusions}
We have introduced an early simple  quantum model of human memory based on the generalized
Grover's algorithm. The output of the model is a probability to find the searched object,
called the "marked item". Es evidenced by Figure 1, such probability  to find the marked item
depends on the following parameters: the number of items $N$, the time to remember (function
of the number of iterations $J$) and the characteristic phase $\phi$.
From another point of view, $\phi$ can be considered a parameter  describing, for fixed $N$
and $J$, the personal emotivity involved in the process. Thus each subject could in principle
use a different phase. However, in our model we hypothesized for simplicity that all the
subjects in the same strategy and emotion condition are described by the same characteristic
phase of the search engine.

The use of Grover's algorithm to describe memory processes has some interesting features: 1)
it involves a clear and well-known mathematical formalism, 2) it hypothesizes the unconscious
knowledge of the marked item, 3) has an iterative nature and 4) the efficiency of the memory
process depends on a simple parameter, the phase, which seems to admit a simple interpretation
in terms of emotivity involved by the items and or by the emotion regulation strategy.

However, we can evidence the following open problems: a) the relation between the time of the
memory task  and the number of iterations $J$ is not known at the moment; b) a precise
mathematical relation between the phases $\phi$ and the emotion regulation strategies is not
known, but only conditions (\ref{condition1},\ref{condition2}); c) it is difficult to estimate
the total number of items $N$ in non-verbal memory tasks.

We conclude by noting that at the moment the use of Grover's algorithm is only an attempt, and
the study of several memory experiments will allow us to decide whether  or not the Grover's
algorithm is a good model for human memory.
\section*{APPENDIX: Mathematical description of Grover's algorithm}
Given a database with $N$ items, we build a complex  Hilbert space $H$ with dimension $N$; the
items are labelled with the integer $i=1,..., N$, and each item $i$ of the database
corresponds to the vector $|i\rangle$. The set of vectors $\{|i\rangle\}$ defines an
orthonormal basis of the Hilbert space $H$, which is called the \textit{computational basis}.
The algorithm requires as the initial state a superposition of the vectors $\{|i\rangle\}$
with equal weights $1/\sqrt{N}$, consistently with equation (\ref{superposition}):
\begin{equation}\label{initial1}
|s\rangle=\sum_{i=1}^{N}\frac{1}{\sqrt{N}}|i\rangle\,.
\end{equation}
Moreover, let there be a unique item $\overline{i}$, that satisfies the condition
$C(\overline{i})=1$,  whereas for all other items $C(i)=0$ (assume that for any state $i$, the
condition $C(i)$ can be evaluated in unit time).  The goal of the quantum algorithm is to
identify the item for which $C(\overline{i}) = 1$, which is performed by repeating
$O(\sqrt{N})$ times the following operation
\begin{equation}\label{grover}
-WI_0 W I_{\overline{i}}\,,
\end{equation}
where (a) $I_{\overline{i}}=I-2|\overline{i}\rangle\langle \overline{i}|$ is the
\textit{selective phase inversion}   (a selective phase rotation of $\pi$ amplitude), which
simply changes the sign to the component $\overline{i}$ and leaves unchanged all the other
components otherwise (where $I$ is the identity matrix), (b)  $W$ is the Walsh-Hadamard
transform defined as $W_{i,j}=(-1)^{i \cdot j}$ where $i,j\in [0,N]$  and $\cdot$ is the
bitwise dot product, (c) $I_0=I-2|0\rangle\langle 0|$ is a \textit{selective phase inversion}
(selective phase rotation   of $\pi$ amplitude), which simply changes the sign to the vector
$|0\rangle$ of the computational basis and leaves unchanged all the other components $i\neq
0$.

The Grover's algorithm is often described as a first selective inversion $I_{\overline{i}}$,
followed by  the operation $-WI_0 W$, called the diffusion transform, which admits the simple
interpretation of inversion about the average:
\begin{equation}\label{inversion}
D=-I+2P,\,\, P_{ij}=\frac{1}{N}
\end{equation}
We now briefly show how the amplitudes (\ref{superposition}) change after $J$ iterations of
operation (\ref{grover}): first we note that the initial superposition (\ref{initial1}) can be
written as
\begin{equation}\label{initial2}
|s\rangle=sin(\beta)|\overline{i}\rangle + cos(\beta)|c\rangle\,,
\end{equation}
where $|c\rangle=(N-1)^{-1/2}\sum_{i\neq \overline{i}}|i\rangle$ and
$\beta=arcsin(1/sqrt{N})$. In the Hilbert space spanned by $|\overline{i}\rangle$ and
$|c\rangle$, we can write the Grover's operator (\ref{grover}) in the matrix form
\begin{equation}\label{grover}
\left[\begin{tabular}{cc}
$cos(2\beta)$ & $sin(2\beta)$\\
$-sin(2\beta)$ & $cos(2\beta)$
\end{tabular}\right]
\end{equation}
After $J$ iterations, the amplitude relevant to vector $|\overline{i}\rangle$ is
$sin[(2J+1)\beta]$. The maximal probability to obtain $\overline{i}$ is thus obtained when
$J=\pi/4\beta -1/2$, which is almost equal to $\pi\sqrt{N}/4$ for large $N$.

An important feature of the Grover's algorithm is that a number of iterations lower or higher
than the  optimal number leads to a lower probability to find the marked item. In the case of
lower iterations, we have the \textit{undercooking}, while in the higher case the
\textit{overcooking}.
%
\subsection*{A simple example: N=4}
We consider the following vector $1/2(1,1,1,1)$ in the standard basis
$|00\rangle,|01\rangle,|10\rangle, |11\rangle$: it represents the initial state on which the
algorithm acts. Let us suppose for example that the state we want to find is $|11\rangle$.
Thus we apply a conditional phase shift, obtaining the vector $1/2(1,1,1,-1)$, which can be
written as
$$
\frac{1}{2}[|0\rangle(|0\rangle+|1\rangle)+|1\rangle(|0\rangle-|1\rangle)]
$$
and evidently represents an entangled state of two qubits (since the conditional phase shift
is a nonlocal operation).

Now we apply the diffusion operator of equation $D$ (\ref{inversion}),  which is the inversion
about the average: since the average of the previous state is $1/4$, the inversion about the
average leads to 0 for all $i\neq \overline{i}$, and 1 otherwise. Thus we obtain the state
$|11\rangle$ after only one iteration, which is a peculiarity of the $N=4$ case.
%
\subsection*{Generalized Grover's algorithm}
A generalized version of Grover's algorithm, introduced by Long et al. \cite{long1}, replaces
the inversion of the marked $|\overline{i}\rangle$ state by an arbitrary phase rotation
$\theta$ and the inversion of the state $|0\rangle$ by an arbitrary phase rotation $\phi$. The
operator used within formula \cite{Grover1} becomes now $-WR_0 W R_{\overline{i}}$, where we
used the selective rotations
\begin{eqnarray}
R_{\overline{i}}=I-(1-e^{i\theta})|\overline{i}\rangle\langle \overline{i}|\\
R_{0}=I-(1-e^{i\phi})|0\rangle\langle 0|
\end{eqnarray}
It has been the shown that in such case different values of $\phi$ and $\theta$ leads to
destroy  the quantum search. On the contrary, the phase matching condition $\phi=\theta$ leads
to an efficient quantum algorithm. In fact, the application $J$ times of the generalized
Grover's operator $-WR_0 W R_{\overline{i}}$ can be written as the matrix
\begin{equation}\label{grover}
\left[\begin{tabular}{cc}
$cos[2Jsin(\theta/2)\beta]$ & $sin[2Jsin(\theta/2)\beta]$\\
$-sin[2Jsin(\theta/2)\beta]$ & $cos[2Jsin(\theta/2)\beta]$
\end{tabular}\right]\,.
\end{equation}
Thus the amplitude relevant to the marked item after $J$ iterations is
$sin[(2J+1)sin(\theta/2)\beta]$, which for $\theta=\pi$ is identical to the standard Grover's
algorithm. This means that values of $\theta$ strictly lower or higher than $\pi$ entail a
lower probability to find the marked item at $J$ fixed, or a higher computational time.
 \footnotesize
%
%

\end{document}